\documentclass[aps,pre,reprint,letterpaper,showpacs,amsmath,amssymb]{revtex4-1}
\usepackage{graphicx}

\newcommand{\w}{\bf}    
\newcommand{\al}{\alpha}
\newcommand{\om}{\omega_c}
\newcommand{\de}{\delta}

\newcommand{\bb}{B_{LS}} 
\newcommand{\ls}{\langle} 
\newcommand{\rs}{\rangle} 
\newcommand{\mean}[1]{\langle #1 \rangle}
\newcommand{\ov}[1]{\overline{#1}}
\newcommand{\RR}{\rho}

\newcommand{\ffrac}[2]{\raisebox{.5pt}{\mbox{\footnotesize$\displaystyle\frac{#1}{#2}$}}}

\newcounter{ourcount}
\begin{document}

\title{Diffusion of charged particles in a stochastic force-free magnetic field}
\date{\today}

\author{Ya.N. \surname{Istomin}}
 \altaffiliation[Also at Moscow Institute of Physics and Technology, \\
Institutskii per. 9, Dolgoprudnyi, 
Moscow region, 141700, Russia]
{}
\author{A.M. \surname{Kiselev}}%
\affiliation{%
P.N.~Lebedev Physical Institute, 
Leninsky Prospect 53, Moscow 119991, Russia.
}%

\begin{abstract}

We study diffusion of charged particles in stationary stochastic magnetic field ${\bf B}$ with zero mean, $ \mean{\bf B} = 0 $. In the case when electric current is carried by electrons, the field is force-free, $\mathrm{curl} \,{\bf B} = \alpha{\bf B} $, where $\alpha({\bf r})$ is an arbitrary scalar function. In a small region where the function $\alpha $ and the field magnitude $|{\bf B}|$ are approximately constant, the equations of motion of charged particles are integrated and reduced to the equation of mathematical pendulum. The transition from trapped to untrapped particles is continuously traced. Averaging over the magnetic field spectrum gives the spatial diffusion coefficient $D$ of particles as a function of the Larmor radius $r_L$ in the large-scale magnetic fields ($B_{LS}$) and magnetic field correlation length $L_0$. The diffusion coefficient turns out to be proportional to the Larmor radius, $D\propto r_L $, for $r_L <L_0 / 2\pi $, and to the Larmor radius squared, $ D \propto r_L^2 $, for $ r_L> L_0 /2\pi $. We apply obtained results to the diffusion of cosmic rays in the Galaxy, which contains a large number of independent regions with parameters $L_0$ and $B_{LS}$ varying in wide range. 
We average over $B_{LS}$ with the Kolmogorov spectrum
and over $L_0$ with the distribution function $f(L_0)\propto L_0^{- 1+ \sigma}$. For
the practically flat spectrum $\sigma = 1/15$, we have $ D\propto r_m^{0.7}$, which is consistent with observations.

\end{abstract}
\maketitle


\section{introduction}
It is generally acknowledged that the propagation of charged particles in the turbulent magnetized plasma has the form of diffusion.
This fact has many astrophysical applications, such as the problem of the cosmic rays (CRs) lifetime in the Galaxy, diffusive shock
acceleration in supernova remnants, and the CRs propagation in molecular clouds. However, to calculate the coefficient $D$
of the CRs spatial diffusion, many authors use approximate phenomenological models.

Depending on the energy particles can be divided into magnetized,
whose Larmor radius $r_L$ is much smaller than the correlation
length $L_0$ of the magnetic field, and non-magnetized. To a
first approximation non-magnetized particle moves along a
straight line, which direction slowly changes. The diffusion
coefficient for non-magnetized particles was calculated
analytically, see for example \cite{Plotnikov11}, where magnetic field is assumed to be
delta-correlated in time. To describe the motion of magnetized
particles is much more difficult. Magnetized particles are divided
into two types: trapped, many times reflecting by magnetic
mirrors, and untrapped.

Now the most developed approach for calculation of diffusion coefficients for magnetized particles are the quasilinear theory (QLT),
proposed by Jokipii \cite{Jokipii66}, and its generalizations.
In the QLT it is assumed that the magnetic field is a sum of mean homogeneous field $H$ and random fluctuations $\delta B$, and $\delta B \ll H$.
So the particle trajectory slightly deviates from trajectory in the mean magnetic field. It allows one to find the pitch angle diffusion
coefficient and estimate the mean free path of a particle.
However, this approach is correct only in case of small fluctuations, $\delta B \ll H$.
But in many astrophysical objects magnetic fluctuations are comparable with mean field or even exceed it, $\delta B > H$. So the problem of particle diffusion in magnetic field with large fluctuations is of great practical importance.

In recent years it have been made attempts to extend QLT to the case of large magnetic fluctuations. Several generalizations of the QLT have been proposed. In the paper \cite{Schalchi09} the second order quasilinear theory was developed.
Similar to the QLT the value of a fluctuating magnetic field is considered to be weak. The particle trajectory is
calculated up to the first order in $\delta B$.
As a result expression for $D$ is calculated to the second order in $\delta B$. Next, the results were
extended to the case of large fluctuations $\delta B > H$ without any changes. 
Whereas the expression for $D$ in the standard QLT 
contains integral of the delta function over the wave numbers $k$, the second order QLT
replaces it by an integral of the Gaussian with the relative width
$\delta B /(H\mu)$, where $\mu$ is cosine of the pitch angle. Clearly in the 
$\delta B \gg H$ limit one deals with the function of a large width.
This transition from the delta-function to the function of a large width is at least unjustified.

Matthaeus et al. \cite{Matt03} proposed another generalization of the QLT, the Nonlinear Guiding Center theory (NLGC).
This theory makes two key assumptions. First, the pair correlation function of particle velocities,
$\mean{v (t) v (t ')}$, is supposed to be known and depends only on a time difference
$t-t'$. Second, a strong assumption about the spatial distribution of particles is made, namely, the function
$\mean{e^{i \vec{k} \vec{r}(t)}}$ is written quite arbitrarily.
This allows obtaining an integral equation for the diffusion coefficient $D_\perp$ transverse the mean magnetic field.
It should be mentioned that NLGC, as well as many other generalizations of QLT, is based on the assumption that particles move
along the magnetic field lines. Actually, the problem of random walk of magnetic field lines rather than the problem of particle diffusion is solved. 
However, in reality, only untrapped particles move along the field lines. Thus the motion of a particle is no more described by these theories, as the particle is trapped.

The main question studied in the aforementioned papers is the anisotropy 
$D_\perp/D_\parallel$ of diffusion coefficients. The dependence of $D_\parallel$ on the particle energy is often assumed to be already known.

Dogiel et al. \cite{DI87} calculated $D$ by a different method. They assumed that on the scale less than the correlation length $L_0$ of the magnetic field a magnetized particle moves along a straight line. That gives the estimate
of the diffusion coefficient $D = v_0 L_0$. However, this assumption is rather rough. This approach also takes into account only untrapped
particles, and possibly gives too high value for the diffusion coefficient.

In some astrophysical applications the Bohm diffusion approximation is used, see for example 
\cite{Zirak12}. The Bohm
diffusion \cite{Bohm}, $D_B = cT/(16eH) =
v_T^2/(16\omega_c)$, is the anomalous diffusion of particles transverse
the mean magnetic field $H$, related to their scattering by a plasma
turbulence. In this case the characteristic time of particle
scattering $\tau$ is of the order of the cyclotron
rotation period $\tau = 16/(3\omega_c)$, rather than
the time between classical collisions. Applying to the
collisionless diffusion in a random magnetic field, the term Bohm
diffusion means $D \propto r_L$.

An alternative way is to calculate particle diffusion in the magnetic field numerically
\cite{Casse02,Plotnikov11}. The magnetic field ${\w B(r)} = {\w
H} + \de {\w B(r)}$ is realized on the discrete grid in space.
Fluctuations $\de{\w B (r)}$ are generated as the sum of plane
waves with random phases, polarizations and directions. They
correspond to a fixed spectrum over wave vectors. The Kolmogorov spectrum is
usually assumed. After that the equations of particle motion
in this magnetic field are solved numerically. The results are
averaged over a large number of particles and magnetic field
realizations.
Authors observe transition from the ballistic regime $\mean{\Delta r^2} \propto v^2 t^2$
to the diffusion regime $\mean{\Delta r^2} \propto D t$ at large times,
and measure the diffusion coefficient. In~\cite{Casse02}
the diffusion coefficient for magnetized particles in the magnetic field with the Kolmogorov spectrum
and the arbitrary ratio $\delta B/H$ was calculated.
The result of QLT was confirmed: $D \propto r_L^{1/3}$. 

{However in recent years numerical calculations of particle diffusion provide us new interesting facts.
First, resolution increased due to growing computer power.
In the recent paper~\cite{Snodin16} the following
approximation to the numerical results was obtained
\begin{equation}
D_\parallel = 0.75 v_0 r_L +
\frac{1}{3} v_0 L \frac{H^2}{\delta B^2}\Bigl( \frac{r_L}{L_0}\Bigr)^{1/3}.
\end{equation}
At zero mean field, $H=0$, this expression gives the Bohm diffusion coefficient
$D \propto r_L$.
It should be noted that in the process of magnetic field generation the
range of wave vectors, in which the sum is taken, usually does not exceed $k_{max}/k_{min} = 256$.
This limits the range of variation of the ratio $r_L/L_0$ by two orders of magnitude.
The result $D \propto r_L$ was confirmed in~\cite{Shalchi14}.

Second, in several articles magnetic field is calculated as a result of direct numerical MHD simulations, 
instead of generation as a given sum of waves.
It started with the paper by Beresnyak et al.~\cite{Beresnyak11}.
In~\cite{Cohet} two turbulence models were used.
In the model with compressible forcing, the results reproduce the QLT.
In the model with solenoidal forcing, the results do not coincide with existing theories. Even the power index varies with the magnetic turbulence level.
In~\cite{Snodin17} magnetic field is calculated as a solution of induction equation with prescribed velocity.
Authors found that the presence of magnetic intermittency changes $D$ significantly. 
Thus there is a question of the applicability of previous results with magnetic field, calculated as a sum of plain waves to real turbulence.
}

In any case, the analytical approach provides answers regardless of numerical schemes
and parameters, and has a great
importance for understanding the physics of diffusion process in a random magnetic field.

The outline of the article is as follows. In Sec.~\ref{sec:general} we write out general equations and reproduce the result for diffusion of non-magnetized particles.  
In Sec.~\ref{sec:smalR} we describe our model of the force-free magnetic field and calculate the diffusion coefficient for magnetized particles. 
In the last Sec.~\ref{sec:Galaxy} we apply our results to cosmic rays propagation in the Galaxy. 

\section{Particle diffusion}\label{sec:general}

We consider a motion of a particle with charge $q$ and velocity $v_0$
in a stationary magnetic field ${\w B}({\w r}) $. We introduce the notations:
$B_0$, such that $B_0^2 = \mean{\w B^2}$, and
${\w b} = {\w B}/B_0 $.
We also introduce the cyclotron frequency of a particle in the magnetic field with magnitude $B_0$,
$\om = qB_0/mc\gamma$, where
$\gamma = (1-v_0^2/c^2)^{- 1/2}$.
The random magnetic field ${\w B({\w r})} $ has zero mean value,
$\mean{\w B} = 0$, so its pair correlation function is isotropic 
$\ls B_i({\w x})B_j({\w x + r})\rs = \Phi_{ij}({|\w r|})$.
It is a smooth function, decreasing at large distances $r$.
The structure of the correlation function $\Phi_{ij}({\w r})$ was investigated in papers dealing with the problem of the magnetic dynamo in a turbulent conducting medium \cite{Schek}, as well as in turbulent weakly ionized gas, from which molecular clouds consist of \cite{IstKis13}.
In the present paper we choose it in the form
\begin{equation}\label{BBcorrelator}
\mean{B_i(0)B_i(r)} = \bb^2 \Bigl[ 1 - \left(\frac{r}{L_0}\right)^{2\beta} \Bigr],  \, r<L_0,
\end{equation}
and $\mean{B_i(0)B_i(r)} =0$ for $r > L_0$.
Here $L_0$ is the correlation length of the magnetic field, $\bb$ is the magnitude of large-scale magnetic field, $\beta$ is the spectral index.
For the Kolmogorov spectrum $\beta = 1/3$. 

The particle motion is described by the equations
\begin{align}\label{E:motioneq}
\frac{d r_i}{dt} &=v_i, \\  \nonumber
\frac{d v_i}{dt} &= \om e_{ikl}v_k b_l.
\end{align}
Here $e_{ikl}$ is the unit antisymmetric tensor. We assume
that the particle motion has diffusive character at large times, i.e.
$\mean{r^2 (t)} = 2Dt$, where brackets $\mean{...}$ denote averaging over random
magnetic field realizations. Consequently,
\begin{equation}\label{D_general}
\mean{D} = \frac{1}{2}\,\frac{d \mean{r^2}}{dt} = \mean{r_i v_i}.
\end{equation}
Note that we deal with correlation functions at coincident moments of time.

\subsection*{Large Larmor radius}
In this subsection we consider non-magnetized particles, i.e. particles with large Larmor radius $r_L > L_0$.
In this case it is known that the diffusion coefficient is proportional to the Larmor radius squared, $\mean{D} \propto r_L^2$, see for example \cite{Plotnikov11}.
We briefly explain this result. 

When a particle with a Larmor radius much greater than $ L_0 $ passes through one region with the size
$ \simeq L_0 $, it slightly deviates from the straight line at a small angle $ \de\phi $.
When the particle travels a large path $ s \gg L_0 $, its motion becomes diffusive. We define $\tau$ the time it takes a particle to deviate at an angle of about $\phi\simeq \pi$. Thus, $ 2D_{\phi} \tau = \pi^2 $, where $ D_{\phi} $ is the angular diffusion coefficient, $ D_{\phi} = d\ls \delta \phi^2 \rs / dt / 2 $. And we have for the spatial diffusion coefficient 
\begin{equation}\label{D_help}
\mean{D}=\frac{1}{3} v_0^2\tau = \frac{\pi^2}{6}\frac{v_0^2}{D_\phi}.
\end{equation}
We define the deviation angle as $ (\de \phi)^2 = \de v_i^2 /v_0^2 $.
To find the angular diffusion coefficient we use the equations of particle motion  (\ref{E:motioneq}).
We write
\begin{equation}
\de v_i(t) = \frac{q}{mc\gamma} \, \int_{0}^t \, dt'  e_{ikl}v_k B_l.
\end{equation}
Hence
\begin{eqnarray}
\hspace{-1cm}
&&\frac{1}{2} \frac{d}{dt} \mean{\de \phi^2} = \\   \nonumber
&&\Bigl(\frac{q}{mc\gamma v_0}\Bigr)^2 \int_{0}^t \, dt' e_{ikl} e_{imn} \mean{v_k(t') B_l(t') v_m(t) B_n(t)}.
\end{eqnarray}
We assume that the particle velocity slightly deviates from the initial velocity, $v_i \approx v_{0i}$, and use the relation
$\mean{v_k v_m} = v_0^2\delta_{km}/3 $. As a result we get
\begin{equation}
\frac{1}{2} \frac{d}{dt} \mean{\de \phi^2} =    
\frac{2}{3v_0} \Bigl(\frac{q}{mc\gamma}\Bigr)^2
\int_0^{L_0} dr \mean{B_i(0)B_i(r)}.
\end{equation}
Using the magnetic pair correlation function (\ref{BBcorrelator}) we have
\begin{equation}\label{D_phi}
D_\phi=\frac{4\beta}{3(2\beta+1)}\frac{v_0L_0}{r_L^2},
\end{equation}
where $r_L$ is a Larmor radius in large-scale magnetic field, $r_L=v_0/\omega_c$ for $B_0=B_{LS}$. 
Now the spatial diffusion coefficient is calculated from~(\ref{D_help}).
Thus, if the random magnetic field is a superposition of
regions with the same correlation length $ L_0 $ and the same magnitude of large-scale magnetic
field $ \bb $, we have for the particle with $ r_L> L_0 / 2\pi $
\begin{equation}\label{DlargeA}
\mean{D} = \frac{\pi^2}{6}\frac{v_0^2}{D_\phi}=\frac{\pi^2(2\beta+1)}{8\beta}\frac{v_0 r_L^2}{L_0}.
\end{equation}

\section{Force-free field}\label{sec:smalR}

\subsection{Particle motion}

It is known that the electric current in a plasma is created by electrons. Then, in the absence of an electric field,
the Ampere force acting on the electrons must be zero due to the low mass of the electrons,
${\w j}\times {\w B} = 0$.
In this case the magnetic field is force-free, $\mathrm{curl} \, {\w B} = \alpha{\w B}$.
Due to
$\nabla \,\mathrm{curl} \, {\w B}=0$, we have
\begin{equation*}
\nabla(\al\,{\w B}) = \al\, \nabla {\w B} + {\w B} \nabla\al = {\w B} \nabla\al  =0,
\end{equation*}
so value of $\alpha({\w r})$ is constant along the magnetic field line, and
{the magnetic field is orthogonal to the vector $\nabla\alpha$. 
We choose the $z$ axis along the direction $ \nabla\alpha $, then $B_z =0$.
Magnetic field lies in the plane $ (x, y) $, which is orthogonal to the axis $ z $ and have components $ B_x, \, B_y $. 
From the equation $ \mathrm{curl} \, {\w B} = \alpha {\w B} $ we obtain relations 
$ \partial B_x / \partial z = \alpha B_y $ and 
$ \partial B_y / \partial z = - \alpha B_x $.
It follows that $ B_x^2 + B_y^2 = B_0^2 = \text{const}(z) $.
The local magnetic field then has the following configuration
\begin{equation}\label{Bwithz}
{\w B(r)}=B_0 \left(\sin\int^z\al(z')dz', \cos\int^z\al(z')dz',0\right).
\end{equation}
This configuration is reproduced when $z$ increased by $z^\star$,
such that $\int_0^{z^\star}\al(z')dz' = 2\pi$. 
Therefore, it is sufficient to consider the motion of charged particles in the region 
$0 < z< z^\star$ only. 
Such a small region we call a magnetic cell in the following.
In one cell the magnetic field lies in the $(x, y)$ plane, but its direction changes with $z$.
We assume that the magnetic field in the whole system is a superposition of individual magnetic cells. 
But different cells have different sizes $z^*$ 
(so different values of $\al$), various directions of $\nabla\alpha$ and different values of the magnetic field magnitude $B_0$. 
First we determine the motion of the charged particle in one magnetic cell. 
In the next subsection we will average over cells with various parameters and orientations. 

The advantage of this approach is that one cell is a trap for
charged particles in the $z$-direction. 
In order to show this we assume that on the interval $(0, z^\star)$ the function $\al(z)$ does not change strongly  $\Delta\al/ \al <1/\pi$. Then, introducing the mean value  ${\bar\al}=2\pi/z^\star$, we replace $\int^z\al(z')dz'$ by ${\bar \al}z$. 
Since we will produce averaging over ${\bar \al}z$, we omit the bar symbol above $\al$ in what follows.

The function of $\alpha(\w r)$ can depend on the transverse coordinates $(x, y)$ also. 
We define $ r_\perp = \sqrt{x^2+y^2}$ and choose the origin $x=y=0$ at the point where $B_z = 0$. Since $\nabla_\perp\alpha = 0 $ at $ r_\perp = 0 $, one can represent $\alpha = \alpha_0 [1\pm(r_\perp / l_\perp)^2] $ near the origin. Here $ l_\perp $ is the characteristic scale of variation of $\alpha$ in the transverse direction. Substituting this expression into equations $ \mathrm{curl}{\bf B} = \alpha {\bf B} $ and $ \nabla {\bf B} = 0 $, we obtain formulas for $ B_z $ and $ \delta B_\perp $:
$$
|B_z|\simeq\frac{1}{3}\frac{\alpha_0 r_\perp^3}{l_\perp^2}B_0, \, |\delta B_\perp|\simeq\frac{1}{12}\frac{\alpha_0^2r_\perp^4}{l_\perp^2}B_0.
$$
We see that the longitudinal magnetic field grows as $B_z \propto r_\perp^3 $. It prevents particles from moving in the transverse direction far from $r_\perp=0$, so $ r_\perp \simeq (v_0 / \omega_c) |B_0 / B_z|$. Thus, we obtain the following estimate for the longitudinal magnetic field in one cell,
$$
\left(\frac{B_z}{B_0}\right)^4 < \left(\frac{v_0\alpha}{2\pi\omega_c}\right)^2 \nabla_\perp \left(\frac{1}{\alpha}\right).
$$
Here we express $ l_\perp $ through $ \nabla_\perp\alpha $. 
If the transverse gradient of the longitudinal scale is not very large,
$$ \nabla_\perp \left( \frac{2 \pi}{\alpha} \right) 
< \left( \frac{2 \pi \omega_c }{v_0 \alpha} \right)^2, $$ 
then the longitudinal magnetic field $ B_z $ is less than transverse one $B_0$. 
Actually, for magnetized particles, the right-hand side of the last inequality is greater than unity.

Thus, for not very large gradients of $\alpha$, one can assume that $B_z=0$. In the following we consider the particle motion in the magnetic field
\begin{equation}\label{B}
{\w B(r)}=B_0 \left(\sin \al z, \cos \al z ,0 \right).
\end{equation}
We substitute the magnetic field (\ref{B}) in the equations of motion (\ref{E:motioneq}) 
and we have
\begin{equation*}
\frac{d{\w v}}{dt}= \om
\begin{pmatrix}
-v_z \cos\alpha z\\ v_z \sin\alpha z\\
v_x \cos\al z - v_y \sin\al z
\end{pmatrix}
\end{equation*}
The first two equations can be easily integrated
\begin{align*}
v_x + \frac{\om}{\al} \sin\al z &= C_1 \\
v_y + \frac{\om}{\al} \cos\al z &= C_2
\end{align*}
So the system is reduced to one equation for $z(t)$,
\begin{equation}
\frac{d v_z}{dt} = - C \om \sin(\al z-\theta_0),
\end{equation}
where we introduce the notations $C = (C_1^2 + C_2^2)^{1/2}$ and $\theta_0$, defined by the equations
$C_1/C=\sin \theta_0$, $C_2/C=\cos \theta_0$.
Introducing the new variable
\begin{equation}
\psi = \al z - \theta_0,
\end{equation}
we reduce (7) to the equation of the mathematical pendulum
\begin{equation}
\frac{d^2\psi}{dt^2} = - \al C \om \sin\psi.
\end{equation}
Integrating this equation we obtain the energy conservation law
\begin{equation*}
\frac{v_z^2}{2} - \frac{C\om}{\al} \cos \psi = \mathrm{const} \stackrel{\mathrm{def}}{=} W.
\end{equation*}
We define the new notation $\kappa^2 = 2C\om/(\al W + C\om)$ and we obtain
\begin{equation}\label{pendulum}
\frac{d \psi}{dt} = \mathrm{sign}(v_z)\frac{2}{\kappa} (\al C \om)^{1/2}\, \left(1-\kappa^2 \sin^2 \frac{\psi}{2}\right)^{1/2}.
\end{equation}
In this expression $\mathrm{sign}(v_z)=1$ at $v_z>0$ and $\mathrm{sign}(v_z)=-1$ at $v_z<0$.
We see that charged particles in a force-free field are divided into two classes: trapped in the $z$-direction (particles with $\kappa >1$)
and untrapped in the $z$-direction (particles with $\kappa < 1$). The trapped particles oscillate between points $\psi = \pm 2\arcsin(1 /\kappa)$.
The characteristic amplitude of velocity oscillations in the $z$-direction is $\delta v_z\simeq 2(\om|v_\parallel|/\alpha)^{1/2}$. Here
$v_\parallel$ is the velocity of the particle in the plane $(x, y)$, where the magnetic field lines lie.
The characteristic amplitude of longitudinal velocity oscillations is $\delta v_\parallel\simeq\om/\al$. In the strong magnetic
field, $\om>\alpha v_0$, fluctuations of velocity component $v_z$ of trapped particles are small, $\delta v_z <\om / \alpha$. It means that the Larmor radius $\delta v_z / \om <1 / \alpha$ is less than the characteristic scale of the
magnetic field inhomogeneity $1/\al$. Therefore, if the magnetic field is large enough trapped 
particles undergo ordinary cyclotron rotation.

In an inhomogeneous magnetic field not all particles undergo a finite motion, some of them have a nonzero mean velocity in $z$-direction. The period of particle motion is not equal to the period of the cyclotron rotation, $2\pi/\om$, though the magnetic field magnitude $B_0$ is constant.
For example for particles on the separatrix $\kappa = 1$, the period is infinite. 
So the inhomogeneity of the magnetic field affects the particle motion.

To calculate the diffusion coefficient according to (\ref{D_general}) we need to find a product 
$\ov{r_i v_i}= \ov{x v_x + y v_y + z v_z}$, and average it over large number
of particles, i.e. over the initial conditions ${\w r}(t_0), {\w v}(t_0)$. 
{We assume that different initial conditions are equally probable.
Here and below bar means averaging over initial conditions of the particle. 
It is different from and independent of the averaging $\mean{\dots}$ over the random magnetic field realizations,
which we will carry out in the next subsection.} 
We write
\begin{align*}
&v_x(t) = C_1 - \frac{\om}{\al} \sin \al z(t), \nonumber  \\	
&x(t) = x_0 + \int_{t_0}^t dt' \, v_x = x_0 + C_1 t   \label{xt}
-\frac{\om}{\al} \int_{t_0}^t dt'  \sin \al z(t').
\end{align*}
These expressions contain initial coordinate $x_0$ and initial 
velocity $v_{x_0}$ of the particle. Because the
$\ov{x_0} = 0 $, the first term does not contribute to the expression for
$\ov{r_i v_i}$.
However, as we will see below, the time for particle to travel a certain distance $z$
correlates with its initial position, so the second term must be taken into account. We have
\begin{multline*}
x v_x + y v_y =
\Bigl( C - \frac{\om}{\al} \cos(\al z(t) - \theta_0) \Bigr) Ct  \\
-\frac{C \om}{\al} \int_{t_0}^t dt' \, \cos (\al z(t')- \theta_0) \\
+(\frac{\om}{\al})^2 \int_{t_0}^t dt' \, \cos (\al z(t)- \al z(t'))
\end{multline*}
After calculating the integral, the right-hand side should be averaged over the initial velocities of the particle.
In the first term, we express the final time $ t $ as
\begin{equation*}
t = \int_{\psi_0}^\psi \, \frac{d\psi'}{\dot{\psi}}  =
\frac{\kappa \, \mathrm{sign}(v_z)}{2 (\al C \om)^{1/2}}
\int_{\psi_0}^\psi \,
\frac{d\psi'}{(1-\kappa^2 \sin^2(\psi'/2))^{1/2}},
\end{equation*}
where $\psi' = \al z(t') - \theta_0$. 
In the second and the last terms we also pass to integration over $ \psi'$. 
We introduce new notations 
\begin{align*}
&a = \frac{\al v_0}{\om},\\
&u = \frac{C}{v_0},
\end{align*}
so $a$ is the ratio of the particle Larmor radius $v_0/\om $ to the characteristic scale of the magnetic field inhomogeneity $ 1/\al $.
As a result, we get
\begin{multline*}
x v_x + y v_y =
\frac{\kappa v_0^2 \, \mathrm{sign}(v_z)}{2 (\al C \om)^{1/2}} \times \\
\Bigl[
( u^2 - \ffrac{u}{a} \cos \psi )  \int_{\psi_0}^\psi \, 
\frac{d\psi'}{(1-\kappa^2 \sin^2 (\psi'/2))^{1/2}} \\
- \ffrac{u}{a}  \int_{\psi_0}^\psi \, 
d\psi' \frac{\cos \psi'}{(1-\kappa^2 \sin^2 (\psi'/2))^{1/2}} \\
+ \ffrac{1}{a^2}  \int_{\psi_0}^\psi \, 
d\psi' \frac{\cos (\psi-\psi')}{(1-\kappa^2 \sin^2 (\psi'/2))^{1/2}} 
\Bigr].
\end{multline*}
One can take the integrals in elliptic functions
\begin{equation*}
\hspace{-1cm}\frac{1}{2} \int^\psi \,
\frac{d\psi'}{(1-\kappa^2 \sin^2 (\psi'/2))^{1/2}} =
F(\psi/2,\kappa),
\end{equation*}
\begin{multline*}
\frac{1}{2} \int^\psi d\psi' \,
\frac{\cos \psi'}{(1-\kappa^2 \sin^2(\psi'/2))^{1/2}} = \\
 \frac{2}{\kappa^2} (E(\psi/2,\kappa) - F(\psi/2,\kappa) ) + F(\psi/2,\kappa), 
\end{multline*}
\begin{multline*}
\frac{1}{2} \int^\psi d\psi' \,
\frac{\cos (\psi' - \psi)}{(1-\kappa^2 \sin^2(\psi'/2))^{1/2}} = \\
\cos \psi \Bigl[ \frac{2}{\kappa^2} (E(\psi/2,\kappa) - F(\psi/2,\kappa) ) + F(\psi/2,\kappa) \Bigr] - \\
 -\frac{2 \sin \psi}{\kappa^2}\left(1-\kappa^2 \sin^2 \frac{\psi}{2}\right)^{1/2}.
\end{multline*}

The force-free magnetic field configuration (\ref{B}) is periodic with respect to coordinate $z$ with 
the period $2\pi/\al$. We assume that a random force-free magnetic field is a random superposition of such
magnetic cells with a size $2\pi/\al$, a random orientation and a random value of $\al$. Therefore,
if the particle passes through one cell we have $\psi - \psi_0 = 2\pi \cdot \mathrm{sign}(v_z)$. Substituting these limits into the previous integrals, and taking into account that
\begin{align*}
&E(\frac{\psi_0}{2} + \pi,\kappa) - E(\frac{\psi_0}{2},\kappa) = 2 E(\kappa); \\
&F(\frac{\psi_0}{2} + \pi,\kappa) - F(\frac{\psi_0}{2},\kappa) = 2 K(\kappa),
\end{align*}
where $ K(\kappa) $ and $ E(\kappa) $ are complete elliptic integrals of the first and second kind respectively, we obtain
\begin{multline}\label{xvEK}
x v_x + y v_y =
\frac{2v_0^2}{\om a^2\sqrt{ua}} \,
(ua - \cos \psi_0) \times \\
 \Bigl[
ua \,(\kappa K(\kappa)) 
 -  \left(\ffrac{2}{\kappa} (E(\kappa) - K(\kappa) ) + \kappa K(\kappa) \right) \Bigr].
\end{multline}
We note that $\mathrm{sign}(v_z) $ does not enter in the answer explicitly, because for $v_z <0$ we change the integration limits.
Expression (\ref{xvEK}) is valid for untrapped particles, $\kappa^2 <1 $. For trapped particles, $\kappa^2>1$, we should
analytically continue the expression (\ref{xvEK}) into the region $\kappa^2>1$. The analytical continuation corresponds to integration over
$\psi$ between turning points
$\sin(\psi/2) = \pm 1/\kappa $. We recall that particles are trapped only in the $ z $ direction,
but freely move in directions $ x $ and $ y $.
As a result, we get
\begin{multline*}
x v_x + y v_y =
\frac{2v_0^2}{\om a^2\sqrt{ua}} \,
(ua - \cos \psi_0) \times \\
 \Bigl[ ua \,G_1(\kappa) -  G_2(\kappa) \Bigr],
\end{multline*}
where
\begin{equation}
G_1(\kappa)  =
\begin{cases}
 \kappa K(\kappa), & \kappa<1,  \\
K(1/\kappa), & \kappa>1,
\end{cases}
\end{equation}
\begin{equation}
G_2(\kappa)  =
\begin{cases}
2(E(\kappa) - K(\kappa) )/\kappa + \kappa K(\kappa) ,
& \kappa<1,  \\
2E(1/\kappa) - K(1/\kappa) ,
& \kappa>1.
\end{cases}
\end{equation}
To calculate $z v_z$, we consider trapped particles. 
Since the particle velocity at the turning points is zero, we have $z v_z=0$.
Alternatively, for the pendulum we have 
\begin{equation*}
\overline{z v_z} \propto \ov{\psi \dot{\psi}} =0.
\end{equation*}
We should analytically continue this result onto untrapped particles, 
and finally we obtain for arbitrary $\kappa$
\begin{equation}\label{DthrF}
\ov{D}(a) = \frac{v_0^2}{\om}\, \ov{F}(a),
\end{equation}
where we introduce
\begin{equation}
F(a, {\w v_0}, z_0) = 
\frac{2}{a^2 \sqrt{ua}} \,
(ua - \cos \psi_0) 
 \Bigl[ ua \,G_1(\kappa) -  G_2(\kappa) \Bigr].
\end{equation}

\begin{figure}
\includegraphics[width=7.5cm]{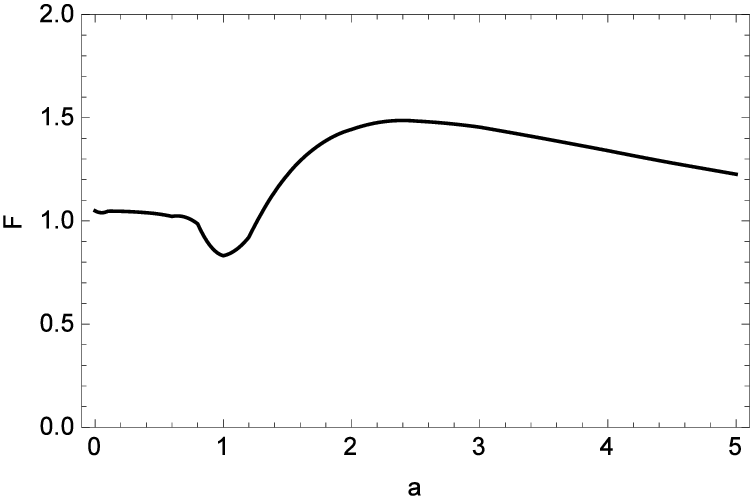}
\caption{Results of numerical calculations of $\ov{F}(a)$. By definition 
$\ov{D}(a)=v_0^2\ov{F}(a)/\omega_c$, see (\ref{DthrF}). The parameter $a$ is dimensionless, 
$a=\alpha v_0/\omega_c$. } 
\label{Fig:Fmean}
\end{figure}

To proceed, we perform the averaging over the direction of initial velocity. We consider
that the velocity magnitude is constant, $|{\w v}| = v_0 $.
Suppose that at the initial moment $ t_0 $ the coordinate $z(t_0) = z_0$ and the particle velocity is
$$
v_{x_0}= v_0 \sin \varphi \sin \delta, \,  v_{y_0} = v_0 \sin \varphi \cos \delta, \,  v_{z_0} = v_0 \cos \varphi.
$$
We introduce the notation $ \tilde{z_0} = \al z_0 - \delta $, and express the
values $ u, \psi_0, \kappa $ through $ \varphi, \delta $ and $ \tilde{z_0} $,
\begin{align*}
&u^2 = \sin^2 \varphi + 2a^{-1}\sin \varphi \cos \tilde{z_0} + a^{-2}, \\
&\cos \psi_0 = \cos (\alpha z_0-\theta_0) = \frac{1}{u} (\sin \varphi \cos \tilde{z_0} + a^{-1}), \\
&\kappa^2 = \frac{4u}{a \cos^2 \varphi + 2u(1 - \cos \psi_0) }.
\end{align*}
We also see that $F(a)$ does not depend on $\al$ except through $a$.
We average $ F(a) $ over $ \varphi, \delta $ with uniform
distribution of initial velocities on the sphere and over $ z_0 $ with a uniform distribution on the interval
$ 0 <z_0 <2\pi/\alpha $,
\begin{equation*}
\ov{F}(a) = \frac{\alpha}{8\pi^2} \int_0^{2\pi/\alpha} \, d z_0 \int_0^{\pi} \, \sin \varphi \, d\varphi
\int_0^{2\pi} \, d \delta\, F(a,\varphi,\tilde{z_0}).
\end{equation*}
Since the parameters $ u, \psi_0, \kappa $ depend only on $ \cos\tilde{z_0} $ and do not depend on $ z_0 $ and $\delta$ separately,
and  the function $\cos\tilde{z_0}$ is periodic, we take the integral over $\delta$ and get
\begin{equation}\label{IntForF}
\ov{F}(a) = \frac{1}{2\pi} \int_0^{\pi/2} \, \sin \varphi \, d\varphi
\int_0^{2\pi} \, d \tilde{z_0} \, F(a,\varphi,\tilde{z_0}).
\end{equation}
We calculate this integral numerically, the results are presented in
Figures \ref{Fig:Fmean}, \ref {Fig:Fmean2}.
We mention that in the limit $ a\to 0 $ the integral (\ref{IntForF}) can be calculated analytically. We have
\begin{equation}\label{F_asimpt}
\ov{F}(a) = \frac{\pi}{3} + o(a), \quad a \to 0.
\end{equation}
From Figure~\ref{Fig:Fmean} we see that the asymptotics (\ref{F_asimpt}) gives a good approximation for the function $\ov{F}(a)$ for $a<1$.
For large values of $a$, $a>3$, the function $\ov{F}(a)$ can be approximated by power law
\begin{equation}\label{F_fit}
\ov{F}(a)  \approx  3.6\, a^{-0.65}
\end{equation}
with high accuracy, see Figure~\ref{Fig:Fmean2}.

\begin{figure}
\includegraphics[width=7.5cm]{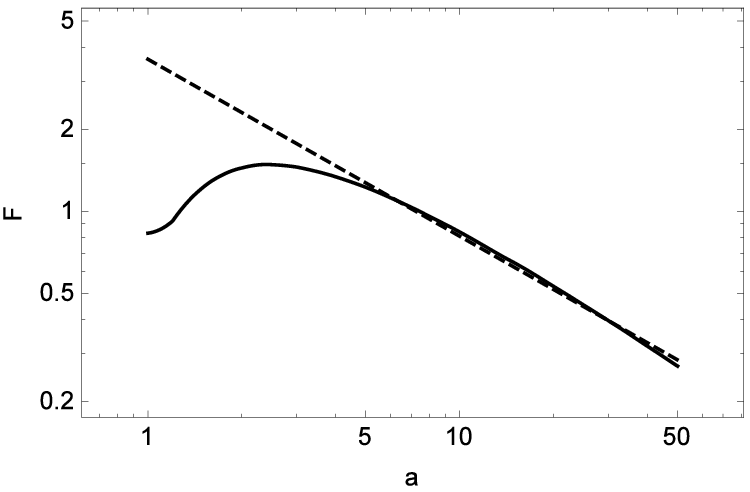}
\caption{Results of numerical calculations of $\ov{F}(a)$, see (\ref{DthrF}), 
for $a>1$ (solid line), logarithmic coordinates.
Dashed line shows the function $\propto a^{-0.65}$.}
\label{Fig:Fmean2}
\end{figure}

\subsection{Averaging over the spectrum}
After averaging over the direction of initial velocity we have the expression for the diffusion coefficient which depends only on the parameters $\al$ and $B_0$ of the magnetic field:
\begin{equation*}
D(\al, B_0) = \frac{v_0^2}{\om} \overline{F(a)},
\end{equation*}
where, we remind,
$\om = q B_0/mc\gamma$, $a=\al v_0/\om$.
Next we average $D$ over the realizations of the random magnetic field, i.e. over the spectrum of magnetic fluctuations $B_0(r)$. 
The spectrum must be consistent with the pair correlation function (\ref{BBcorrelator}), so we  choose it to be a power law
\begin{equation}\label{BSpectrum}
B_0(r) = \bb \left(\frac{r}{L_0}\right)^\beta.
\end{equation}
Actually expression~(\ref{BSpectrum}) describes the root mean square of magnetic field fluctuations of  scale $r$.
Here $\bb$ is the magnitude of large-scale magnetic field and $L_0$ is the correlation length. For the Kolmogorov spectrum \mbox{$\beta = 1/3$}. 
To average $D$ over the spectrum we use the following model. 
We suppose that the scale $r$ of magnetic field variation is equal to the size of the magnetic cell,
so we choose $\al = 2\pi/r$ and have a consistent formula
\begin{equation}\label{22}
\mean{D} = \frac{1}{L_0} \int_0^{L_0} dr \, D\left(\al = \frac{2\pi}{r}, B_0 = B_0(r)\right).
\end{equation}
\begin{figure}
\includegraphics[width=7.5cm]{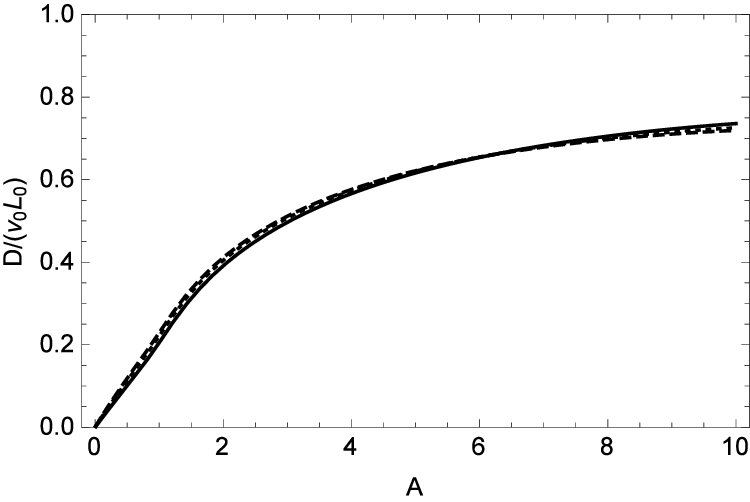}
\caption{Numerical calculations of $\mean{D(A)}$ (\ref{IntForD}) for $\beta=0.2 $ (solid line), $\beta=0.33$ (dotted line), $\beta=0.4$ (dashed line). The parameter $A$ is dimensionless,
 $A=2\pi r_L/L_0$.}
\label{Fig:Dnumeric1}
\end{figure}
Here and below, the angle brackets $ \mean{...} $ mean averaging over the spectrum.
We introduce the notations
\begin{equation*}
\omega_0=\frac{e \bb}{mc\gamma},
\qquad r_L = \frac{v_0}{\omega_0}
\end{equation*}
for the cyclotron frequency $\omega_0$ and the Larmor radius $r_L$ in the large-scale magnetic field $\bb$. We also introduce the parameter
\begin{equation*}
A = \frac{2\pi r_L}{L_0}.
\end{equation*}
Then, the averaging over the spectrum is given by
\begin{equation*}
\ls D \rs = \frac{v_0^2}{L_0} \int_0^{L_0} dr \,
\frac{\overline{F}(a(r))}{\om(r)}.
\end{equation*}
For the choosen spectrum (\ref{BSpectrum}) $\om(r) = \omega_0
(r/L_0)^\beta$, $a(r) = A (r/L_0)^{-\beta-1}$, and we have
\begin{equation}
\mean{D(A)} = \frac{v_0 A}{2\pi} \int_0^{L_0} \, \ov{F}(a)\left(\frac{r}{L_0}\right)^{-\beta} \,dr.
\end{equation}
Changing the variable in the integral we finally obtain
\begin{equation}\label{IntForD}
\mean{D(A)} = \frac{v_0 L_0}{2\pi (\beta+1)} \int_A^\infty \, \ov{F}(a)
\left(\frac{A}{a}\right)^{\xi} \,da,
\end{equation}
where the index is $\xi = 2/(\beta + 1)$. For the fixed value of $\beta$ the result depends only on the parameter $A$.

Using the asymptotics (\ref{F_asimpt}) we treat this integral analitically 
for $A\ll 1$ and get
\begin{equation}\label{D_analitic}
\mean{D(A)} \simeq \frac{A}{6(1-\beta)} v_0 L_0.
\end{equation}
We also calculated the integral (\ref{IntForD}) numerically for different values of $ \beta $, using the previously calculated function $ \ov{F}(a) $.
The result is shown in Figure~\ref{Fig:Dnumeric1}.
We see that the average diffusion coefficient weakly depends on $\beta$.
So we consider only the Kolmogorov spectrum $ \beta = 1/3 $ in what follows.
From Figure~\ref{Fig:Dnumeric3} one can see that
$\mean{D}\propto A$ for $A <1 $. This is consistent with the analytical result (\ref{D_analitic}) and gives the Bohm diffusion coefficient, $ D\propto v_0 r_L $.

For $ A > 1 $, Figure~\ref{Fig:Dnumeric1} shows a slow growth
of $ \mean{D(A)} $.
But as was shown in previous section, the diffusion coefficient for particles with $ r_L > L_0 $ increases proportionally to the square of the Larmor radius, i.e. $\mean{D} \propto A^{2} $.

The reason for the discrepancy is the following. For correct description of diffusion of particles with $ r_L> L_0 $, it is necessary to consider the particle motion on the scale much larger than $ L_0 $,
which contradicts the expression~(\ref{BSpectrum}). Hence the results calculated 
from~(\ref{IntForD}) are applicable only for particles with $ r_L<L_0$.

\begin{figure}
\includegraphics[width=7.5cm]{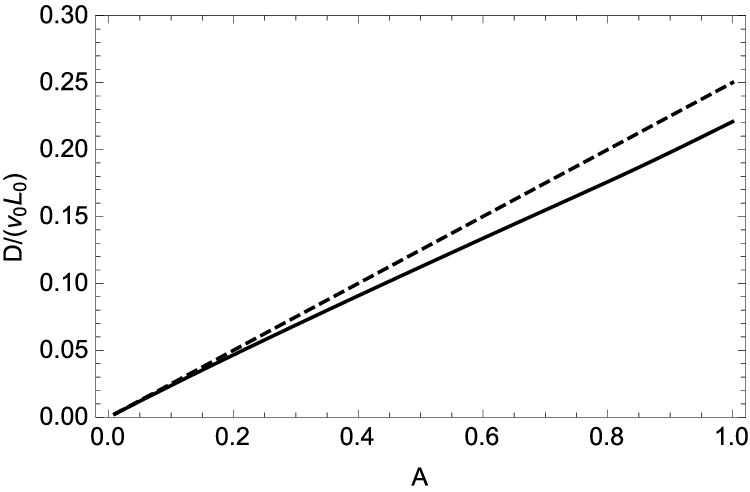}
\caption{Numerical calculations of $\mean{D(A)}$ (\ref{IntForD}) for $\beta=0.33$ (solid line). Dashed line shows the linear function $0.25 \, A$ (see the expression (\ref{preFinal})).}
\label{Fig:Dnumeric3}
\end{figure}

Adopting for the case $A>1$ the result~(\ref{DlargeA}), we approximately represent the diffusion coefficient for the Kolmogorov spectrum
$\beta = 1/3$ as
\begin{equation}
\mean{D} \simeq
\begin{cases}
(1/4) A\, v_0 L_0,  & A<1 \label{preFinal} \\
(5/32) A^2 v_0 L_0,  & A>1.
\end{cases}
\end{equation}
We note that two cases practically coincide at the point $ A = 1 $. One can rewrite this formula in terms of $ r_L $:
\begin{equation}\label{Final}
\mean{D} \simeq
\begin{cases}
(\pi/2) v_0 r_L,  & r_L<L_0/2\pi  \\
(5\pi^2/8)(v_0 r_L^2/L_0),  & r_L>L_0/2\pi.
\end{cases}
\end{equation}
We see that diffusion of particles is primarily determined by their motion in a large-scale magnetic
field and weakly depends on the spectrum of
magnetic fluctuations. The index $ \beta $ of the spectrum (\ref{BSpectrum}) enters only in the common factor in the function
$\mean{D(r_L)}$.
Since the Larmor radius of the particle is proportional to its rigidity ${\cal R}$, $ r_L = {\cal R} / B_{LS} $,
then for small rigidities the diffusion coefficient is proportional to the rigidity (energy), $ D \propto {\cal R} $,
and for large rigidities it proportional to the square of the rigidity, 
$ D \propto {\cal R}^2 $. The result (\ref{Final}) has a simple
physical interpretation. The diffusion coefficient of particles is determined by their motion in a large-scale field (its amplitude is maximal for the spectral index $\beta> 0$) and it is the increasing function of the Larmor radius.

\section{Diffusion of cosmic rays in the Galaxy}\label{sec:Galaxy}

The structure of the magnetic field in the Galaxy is much more complicated than represented in the 
equation~(\ref{BSpectrum}).
{First of all there is nonzero mean magnetic field $\mean{B} \ne 0$ in some parts of the Galaxy. 
It's likely essential in the spiral arms. 
However, the observed high isotropy of cosmic rays \cite{Ginzburg} indicates
that regions with strong mean magnetic field do not have significant impact on the cosmic rays propagation in the whole Galaxy.
Below we discuss the motion of charged particles in a purely random magnetic field. 
We also consider the magnetic field to be force-free.
This assumptrion does not greatly limit the application of our analytical results for the cosmic rays diffusion in the Galaxy. 

The Galaxy consists of a large number of regions, and magnetic fields in different regions are not correlated. 
We assume that inside each region the turbulent magnetic field has the Kolmogorov spectrum (\ref{BSpectrum}) with $\beta=1/3$, but the parameters
$\bb$ and $L_0$ are different in different regions.
It is known that the large-scale magnetic field varies in a wide range, $10^{-6} \, \text{G} \leq B_{LS} \leq 10^{-4} \text{G}$.
Besides the region size $L_0$ also varies in a wide range, from
$\sim 10^{12}$~cm
to the size $\sim 100$~pc of large molecular clouds or clouds of ionized  gas. Therefore it is necessary
to average particle motion over different regions having different values of the correlation length $L_0$ and magnitudes of
large-scale field $B_{LS}$. Let the value of $ L_0 $ varies from zero to the maximum scale $ L_m $.
We denote by $f(L_0)$ the distribution function of sizes $L_0$ of the magnetic field regions,
\begin{equation}
dN = f(L_0) \, dL_0,
\end{equation}
where $dN$ is the number of regions with size $L_0$.
Suppose that $f(L_0)$ is a decreasing power function
(it is normalized to unity),
\begin{equation}
f(L_0)=\sigma L_m^{-1}\left(\frac{L_0}{L_m}\right)^{\sigma-1}, \, 0<\sigma<1.
\end{equation}
For $\sigma \ll 1 $ we have a practically flat spectrum: the number of objects in the range
$ 0 <L_0 <L_1 $ equals $N(L_1) = (L_1 / L_m)^\sigma \simeq \text{const} (L_1) $.
It is naturally to extend the dependence (\ref{BSpectrum}) on large scales when choosing a distribution of magnetic field amplitudes
\begin{equation}
\bb = B_m\left(\frac{L_0}{L_m}\right)^\beta.
\end{equation}
Here $L_m $ is the maximum scale, $ B_m $ is the maximum amplitude of the magnetic field. In regions with different amplitudes of the magnetic
field the particle has a different Larmor radius.
The distribution over the Larmor radii corresponds to the distribution over the magnetic
field, $r_L = r_m (L_0 / L_m)^{-\beta}$, where $r_m$ is the Larmor radius of the particle in the field $B_m$,
$ r_m = v_0 mc \gamma / q B_m $. We introduce the magnetization parameter of the particle in the magnetic field $B_m$,
\begin{equation}
\rho = 2\pi \frac{r_m}{L_m}.
\end{equation}
We will focus on particles with $\RR<1$, i.e. magnetized in the maximum magnetic field.
The average diffusion coefficient is a sum 
$\mean{D} = D_1 + D_2$, where $D_1$ and $D_2$ are the diffusion coefficients averaged over regions where the particle is non-magnetized and magnetized correspondingly. 
Indeed, $D\simeq v_0 \lambda$, so when the particle moves through different regions with different 
average mean free paths $\lambda_1$ and $\lambda_2$, the total mean free path is equal to their sum,
$\lambda = \lambda_1 + \lambda_1$, and similarly for the diffusion coefficients.

The particle is magnetized in some region if \mbox{$r_L<L_0/2\pi$}.
This condition is fulfilled in regions with size $L_0 > L_{cr}$, where
\begin{equation}
L_{cr} = L_m \rho^{\frac{1}{1+\beta}}.
\end{equation}
If $L_0<L_{cr}$ particles are non-magnetized, and the diffusion coefficient is determined by the angular diffusion
\begin{equation}
D_1 = \frac{\pi^2}{6} \frac{v_0^2}{\mean{D_\phi}},
\end{equation}
where we must substitute the value of $D_\phi$ averaged over different scales $L_0$
\begin{multline*}
\mean{D_\phi} = \int_{0}^{L_{cr}} dL_0 \, f(L_0) D_\phi(L_0) = \\
\int_{0}^{L_{cr}} dL_0 \, \frac{\sigma L_0^{\sigma-1}}{L_m^\sigma}\,
 \frac{4\beta}{3(2\beta+1)} \frac{v_0 L_0}{r_m^2} \, \Bigl(\frac{L_0}{L_m}\Bigr)^{2\beta} = \\
\frac{16\pi^2}{3}\frac{\sigma \beta}{(2\beta+1)(2\beta+\sigma+1)} \frac{v_0}{L_m}
\RR^{\frac{\sigma-1}{1+\beta}}
\end{multline*}
Then
\begin{equation}
D_1 = \frac{(2\beta+\sigma+1)(2\beta+1)}{32\sigma\beta} v_0 L_m
\RR^{\frac{1-\sigma}{1+\beta}}.
\end{equation}
If $ L_0>L_{cr}$ particles are magnetized, and
$$D (L_0) = \frac{\pi}{3(1-\beta)} v_0 r_L.$$
Using the dependence $r_L(L_0)$ we calculate
\begin{multline*}
D_2 = \int_{L_{cr}}^{L_m} dL_0 \, f(L_0) D(L_0) = \\
\frac{\pi}{3(1-\beta)} \sigma v_0 r_m \int_{L_{cr}}^{L_m} dL_0 \,  \frac{L_0^{\sigma-1}}{L_m^\sigma}\,\, \Bigl(\frac{L_m}{L_0}\Bigr)^{\beta} = \\
\frac{\sigma}{6(1-\beta)(\beta-\sigma)} v_0 L_m
\left(\RR^{\frac{1+\sigma}{1+\beta}} - \RR \right)
\end{multline*}
Thus, the average diffusion coefficient for particles with $ \rho < 1 $ is
\begin{multline}\label{D_R}
\mean{D} =  v_0 L_m
 \Bigl[ \frac{(2\beta+\sigma+1)(2\beta+1)}{32\sigma\beta} \RR^{\frac{1-\sigma}{1+\beta}} \\
+\frac{\sigma}{6(1-\beta)(\beta-\sigma)}
 (\RR^{\frac{1+\sigma}{1+\beta}} - \RR) \Bigr].
\end{multline}

For small values of the index $ \sigma $ the second term is negligible, so the diffusion coefficient depends on the magnetic rigidity as
$ D \propto {\cal R}^{(1- \sigma) / (1+ \beta)} $.
Since we consider $\sigma>0$, the index
$(1-\sigma)/(1+\beta)<3/4$ for $\beta=1/3$.
So, for the Kolmogorov spectrum $\beta = 1/3$ and
$\sigma = 1/15$ the value of the index is $ 0.7 $, which corresponds to observations \cite{Ginzburg}.

The function $\mean{D (\rho)}$ (\ref{D_R}) for $ \beta = 1/3, \, \sigma = 1 /15$ is plotted in the figure \ref{diff1}.
One can see that it hardly differs from the law
$\mean{D} \propto r_m^{(1-\sigma)/(1+\beta)}=r_m^{0.7}$.
Thus, regions of relatively small sizes, $L_0 <L_{cr}$, where particles are non-magnetized, 
give the main contribution to spatial diffusion of particles with $\rho <1$. 
Large regions, $ L_0> L_{cr}$, where particles spend large time, 
give the contribution to diffusion at least
\begin{equation*}
\frac{16 \sigma^2 \beta}{3(1-\beta)(2 \beta + \sigma + 1) (2\beta + 1)(\beta-\sigma)} \simeq 10^{-2}
\end{equation*}
times less.
For $L_m \simeq 100$~pc, $ B_m \simeq 10^{-4}$~G, $ \beta = 1/3 $ and
$ \sigma = 1/15 $ the diffusion coefficient (\ref{D_R}) for particles with energy 
$\simeq 10$~GeV
is of the order of $1.5 \times 10^{28}\, \text{cm}^2 / \text{s} $, which is in agreement with observations \cite{Ginzburg}.

As for particles with large Larmor radius, $ r_m>L_m / (2\pi)$, the diffusion coefficient for them is
proportional to the square of the magnetic rigidity (which is proportional to its energy)
\begin{equation}
\mean{D} = \frac{(2\beta+\sigma+1)(2\beta+1)}{32\sigma\beta} v_0 L_m\RR^2.
\end{equation}
However, for $ L_m \simeq 100$~pc and $B_m \simeq 10^{-4}$~G condition ${r_m>L_m/(2\pi)}$ corresponds to protons with energies ${{\cal E}> 10^{16}}$~eV.
Particles with such energies are poorly trapped in the Galaxy disk and fill the halo.

\begin{figure}
\includegraphics[width=7.5cm]{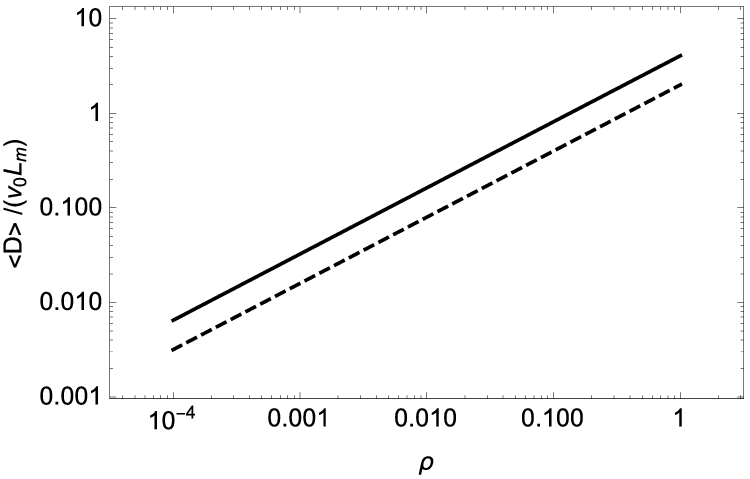}
\caption{The diffusion coefficient
$\mean{D(\rho)}$ (\ref{D_R}) for $\beta=1/3$,
$\sigma=1/15$ (solid line).
Dashed line shows the function $\propto\rho^{0.7}$.} \label{diff1}
\end{figure}

\section{Discussion}
We have considered the motion of charged particles in a random force-free magnetic field,
$\mathrm{curl}{\bf B} = \alpha {\bf B}$.
Such random field can be represented as a superposition of the different magnetic cells. In each cell magnitude of the magnetic field is constant.
The direction of the magnetic field is parallel to one plane, but the direction changes with the coordinate perpendicular to this plane.
The characteristic scale of field variation is $l = 2\pi/\alpha$.

The character of the particle motion in arbitrary magnetic field is
known only in two cases: 1) when $r_L \ll l$ and 2) when $r_L \gg
l$. In the first case the particle motion consists of motion along
the magnetic field, cyclotron rotation and drift in the transverse
direction (gradient drift and curvature drift). In the opposite limit
the particle moves almost in a straight line and
slightly deviates in the direction perpendicular to the magnetic
field.

In our model of the force-free magnetic field the particle motion in one magnetic cell
can be exactly described even in the case when the Larmor radius $r_L$ is of the order of  $l$.
In this case particle performs nonlinear oscillations in the direction perpendicular to the plane in which
a magnetic field line lies.
Some particles (trapped particles) perform the finite motion transforming to the cyclotron rotation for $\kappa\gg1$.
The period of their motion is equal to the period of the cyclotron rotation $2\pi/\omega_c$
only in the limit $r_L \ll l$.
The period of motion increases as the ratio $r_L/l$ approaches unity.
On the separatrix, $\kappa=1$, (see the equation (\ref{pendulum})) the period is infinite. Further,
for $\kappa<1$ the particle freely moves in the direction of the field gradient (untrapped particles).
Thus, we continuously trace a transition from magnetized to free motion.

We found the contribution of the motion of particles in a single magnetic cell in their diffusion coefficient $D$ by averaging $D$ over the initial coordinates and velocities of particles.
We consider initial velocities to be equally probable.
The answer depends only on the dimensionless
parameter $a = \alpha v_0 / \omega_c$, which is the ratio of the cyclotron radius to the length $l$ of the field inhomogeneity. 

Next, we average the diffusion coefficient  over the spectrum of the magnetic field, assuming it to have a power 
form (\ref{BSpectrum}). It turns out, that the
diffusion of particles is mainly determined by the Larmor
radius $r_L$ of particle rotation in the large-scale field and its maximum
scale $L_0$. The spectral index $\beta$ enters only in the factor before the diffusion coefficient.
We find that diffusion coefficient is proportional to the Larmor radius of the particle,
$r_L$, for $r_L <L_0 / 2\pi$ for all reasonable spectral indices
$\beta$, and decreases with the Larmor radius for $r_L> L_0 /
2\pi$. However, for particles with a large Larmor radius $r_L > L_0 / 2\pi$ this approach does not take into account that
particle motion acquires diffusion character only when particle travels a long path $s \gg L_0$. For such particles the diffusion
coefficient can be calculated in the framework of another approach.
Mean free path is determined by the angular diffusion coefficient
$D_{\phi}$ (\ref{D_phi}), and the spatial diffusion coefficient is proportional to
$r_L^2$. Thus, if a random magnetic field is a superposition of regions
with the same size $L_0$ and the same value of large-scale magnetic field
$B_{LS}$, the diffusion coefficient of particles is presented by the formula (\ref{Final}).
It is linear, then quadratic dependence on Larmor radius in the large scale field.

We also discuss the motion of cosmic rays in the Galaxy, where the parameters $B_{LS}$ and $L_0$ of different regions
vary in the wide ranges. Therefore we must average the motion of a particle over different regions. 
We assume that the large-scale 
field distribution has the same spectral index as the small-scale field (\ref{BSpectrum}).
We suppose that the distribution function over
scales also has a power law form, $f(L_0) \propto L_0^{-1+\sigma}, \, \sigma<1$.
After averaging we found that the diffusion coefficient depends on the particle Larmor
radius as $D \propto r_L^{(1-\sigma) / (1+\beta)}$. For $\beta=1/3$ (Kolmogorov spectrum) and $\sigma = 1/15$ (the distribution over scales is practically flat, also called the Harrison-Zel'dovich spectrum 
\cite{Harrison,Zel'dovich} we obtain the spectral index $D\propto r_L^{0.7}$, which consistent with observations of cosmic rays diffusion in the Galaxy.

\section*{Aknowlegements}
We thank D.~Chernyshov for useful discussions. We grateful to the referee for instructive comments, which helped us to find a new straightforward way to get an answer.
This work was done under support of the Russian Foundation for Basic
Research (grant numbers 17-02-00788).

\end{document}